\documentclass[12pt,preprint]{aastex}

\shorttitle{Stellar Oscillator NPOI Observations}
\shortauthors{Baines et al.}

\begin{document}

\title{NPOI Measurements of Ten Stellar Oscillators}

\author{Ellyn K. Baines, J. Thomas Armstrong, Henrique R. Schmitt}
\affil{Remote Sensing Division, Naval Research Laboratory, 4555 Overlook Avenue SW, \\ Washington, DC 20375}
\email{ellyn.baines@nrl.navy.mil}

\author{James A. Benson, R. T. Zavala}
\affil{U.S. Naval Observatory, Flagstaff Station, AZ 86001}

\author{Gerard T. van Belle}
\affil{Lowell Observatory, Flagstaff, AZ 86001} 

\begin{abstract}

Using the Navy Precision Optical Interferometer, we measured the angular diameters of 10 stars that have previously measured solar-like oscillations. Our sample covered a range of evolutionary stages but focused on evolved subgiant and giant stars. We combined our angular diameters with \emph{Hipparcos} parallaxes to determine the stars' physical radii, and used photometry from the literature to calculate their bolometric fluxes, luminosities, and effective temperatures. We then used our results to test the scaling relations used by asteroseismology groups to calculate radii and found good agreement between the radii measured here and the radii predicted by stellar oscillation studies. The precision of the relations is not as well constrained for giant stars as it is for less evolved stars.

\end{abstract}

\keywords{visible: stars, stars: fundamental parameters, techniques: interferometric)}


\section{Introduction} 

Asteroseismology, the study of stellar oscillations, is a powerful tool to infer information about stellar structure with minimal model dependence \citep[see, e.g.,][]{1994ARAandA..32...37B, 2004SoPh..220..137C}. The frequencies of the observed oscillations depend on the sound speed inside the star, which in turn is dependent on properties of the interior such as density, temperature, and gas motion \citep{2010AandA...509A..73C}. The number of stars observed using asteroseismology and the quality of the data have increased dramatically in recent years, thanks to the photometric space missions \emph{MOST} \citep[\emph{Microvariability and Oscillations of STars,}][]{2003PASP..115.1023W}, \emph{CoRoT} \citep[\emph{Convection, Rotation, and planetary Transits,}][]{2006ESASP.624E..34B, 2009AandA...506..411A}, and \emph{Kepler} \citep{2010Sci...327..977B, 2010ApJ...713L..79K}. The resulting stellar parameters are key to the statistical analysis of fundamental stellar properties and for testing stellar interior and evolutionary models \citep[see, e.g.,][]{2011Sci...332..213C}. 

Interferometry has the potential to make important contributions to asteroseismology, in part through the determinations of the targets' sizes \citep{2007AandARv..14..217C}. Using interferometry, we can measure the angular diameters of stars with resolutions down to tenths of a milliarcsecond \citep[see, e.g.,][]{2012MNRAS.423L..16H, 2012ApJ...761...57B}. Once we know the apparent diameter of a star as well as its distance from parallax measurements, we can calculate its physical size. Then we can test the relationships used to derive stellar properties from asteroseismology observations by comparing the radii estimated using the asteroseismology relations to those measured interferometrically.

\citet{2012ApJ...760...32H} presented interferometric diameters of 10 stars that had oscillation measurements from \emph{CoRoT} and \emph{Kepler}. They found an agreement between asteroseismic and interferometric radii of $\lesssim$ 4$\%$ for dwarf stars and $\sim$ 13$\%$ for giant stars. Their sample included five dwarf, one subgiant, and four giant stars. Here we focus on the more evolved stars: one dwarf, four subgiant, and five giant stars.

We observed these stars using the Navy Precision Optical Interferometer (NPOI) in order to measure their angular diameters. We then calculated their radii and effective temperatures, and used spectral energy distribution fits to determine their bolometric fluxes and luminosities. Section 2 discusses the NPOI and our observing process; Section 3 describes the visibility measurements and how we calculated various stellar parameters; Section 4 explores the relationship between radii determined using asteroseismology observations and radii measured interferometrically; and Section 5 summarizes our findings.


\section{Interferometric Observations}

The NPOI is an interferometer located on Anderson Mesa, AZ, and consists of two nested arrays: the four stations of the astrometric array (AC, AE, AW, and AN, which stand for astrometric center, east, west, and north, respectively) and the six stations of the imaging array, of which two stations are currently in operation (E6 and W7) and three more will be coming online in the near future (E7, E10, and W10). The current baselines, i.e., the distances between the stations, range from 16 to 79 m, and our maximum baseline will be 432 m when the E10 and W10 stations are completed within the next year. We use a 12-cm region of the 50-cm siderostats and observe in 16 spectral channels spanning 550 to 850 nm simultaneously \citep{1998ApJ...496..550A}.

Each observation consisted of a 30--second coherent (on the fringe) scan in which the fringe contrast was measured every 2 ms, paired with an incoherent (off the fringe) scan used to estimate the additive bias affecting the visibility measurements \citep{2003AJ....125.2630H}. Scans were taken on five baselines simultaneously. Each coherent scan was averaged to 1--second data points, and then to a single 30--second average. The dispersion of 1--second points provided an estimate of the internal uncertainties.

The target list was derived from the sample of stars with stellar oscillations that were bright enough to observing using the NPOI, which has a magnitude limit of $V = 6.5$. They also had to be resolved with the longest existing baseline, which gives a resolution limit of approximately 1 milliarcsecond (mas). This resulted in a list of 10 targets with stellar oscillation observations available to observe using the NPOI.

We interleaved data scans of the 10 asteroseismic targets with one to three calibrator stars for each target. Our calibrators are stars that are significantly less resolved on the baselines used than the targets. This meant that uncertainties in the calibrator's diameter did not affect the target's diameter calculation as much as if the calibrator star had a substantial angular size on the sky. The calibrator and target scans were measured as close in time and space as possible, which allowed us to convert instrumental target and calibrator visibilities to calibrated visibilities for the target. Preference was given to calibrators within 10$^\circ$ of the target stars, as was the case for 13 of the 16 calibrator stars used. On rare occasions, no suitable calibrator stars were within that angular distance so we resorted to stars that were more distant, with a maximum separation of 17$^\circ$.

We estimated the calibrator stars' sizes by constructing their spectral energy distribution (SED) fits using photometric values published in \citet{1965ArA.....3..439L}, \citet{1981PDAO...15..439M}, \citet{1993AandAS..102...89O}, \citet{1990VilOB..85...50J}, \citet{1972VA.....14...13G}, \citet{1970AandAS....1..199H}, \citet{1991TrSht..63....1K}, \citet{1968tcpn.book.....E}, \citet{1966CoLPL...4...99J}, \citet{2003tmc..book.....C}, and \citet{1993cio..book.....G} as well as spectrophotometry from \citet{1983TrSht..53...50G}, \citet{1998yCat.3207....0G}, and \citet{1997yCat.3202....0K} obtained via the interface created by \citet{1997AandAS..124..349M}. The assigned uncertainties for the 2MASS infrared measurements are as reported in \citet{2003tmc..book.....C}, and an uncertainty of 0.05 mag was assigned to the optical measurements. We determined the best fit stellar spectral template to the photometry from the flux-calibrated stellar spectral atlas of \citet{1998PASP..110..863P} using the $\chi^2$ minimization technique \citep{1992nrca.book.....P, 2003psa..book.....W}. The resulting calibrator angular diameter estimates are listed in Table \ref{observations}.


\section{Results}

\subsection{Angular Diameter Measurement}

Interferometric diameter measurements use $V^2$, the square of the fringe visibility. For a point source, $V^2$ is unity, while for a uniformly-illuminated disk, $V^2 = [2 J_1(x) / x]^2$, where $J_1$ is the Bessel function of the first order, $x = \pi B \theta_{\rm UD} \lambda^{-1}$, $B$ is the projected baseline toward the star's position, $\theta_{\rm UD}$ is the apparent uniform disk angular diameter of the star, and $\lambda$ is the effective wavelength of the observation \citep{1992ARAandA..30..457S}. $\theta_{\rm UD}$ results are listed in Table \ref{results}. Our data files in OIFITS format are available upon request.

\begin{figure}[h]
\includegraphics[width=1.0\textwidth]{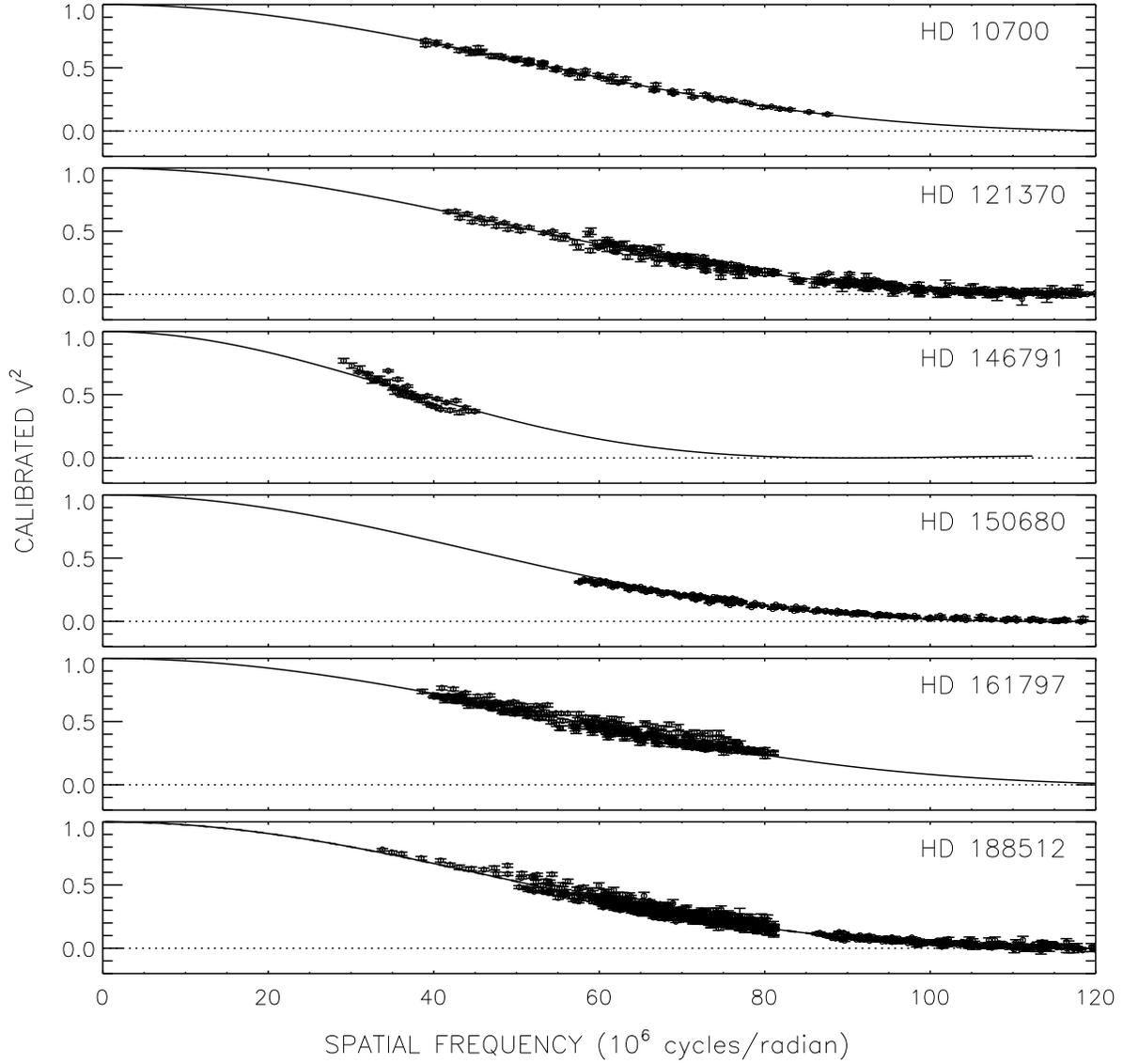}
\caption{$\theta_{\rm LD}$ fits for stars observed with one calibrator. The solid lines represent the theoretical visibility curve for the best fit $\theta_{\rm LD}$, the points are the calibrated visibilities, and the vertical lines are the measurement uncertainties. The uncertainty in the $\theta_{\rm LD}$ fit is not shown because it largely indistinguishable from the best fit $\theta_{\rm LD}$ curve on this scale. See Table \ref{results} for the uncertainty.}
  \label{ldplot1}
\end{figure}

\begin{figure}[h]
\includegraphics[width=1.0\textwidth]{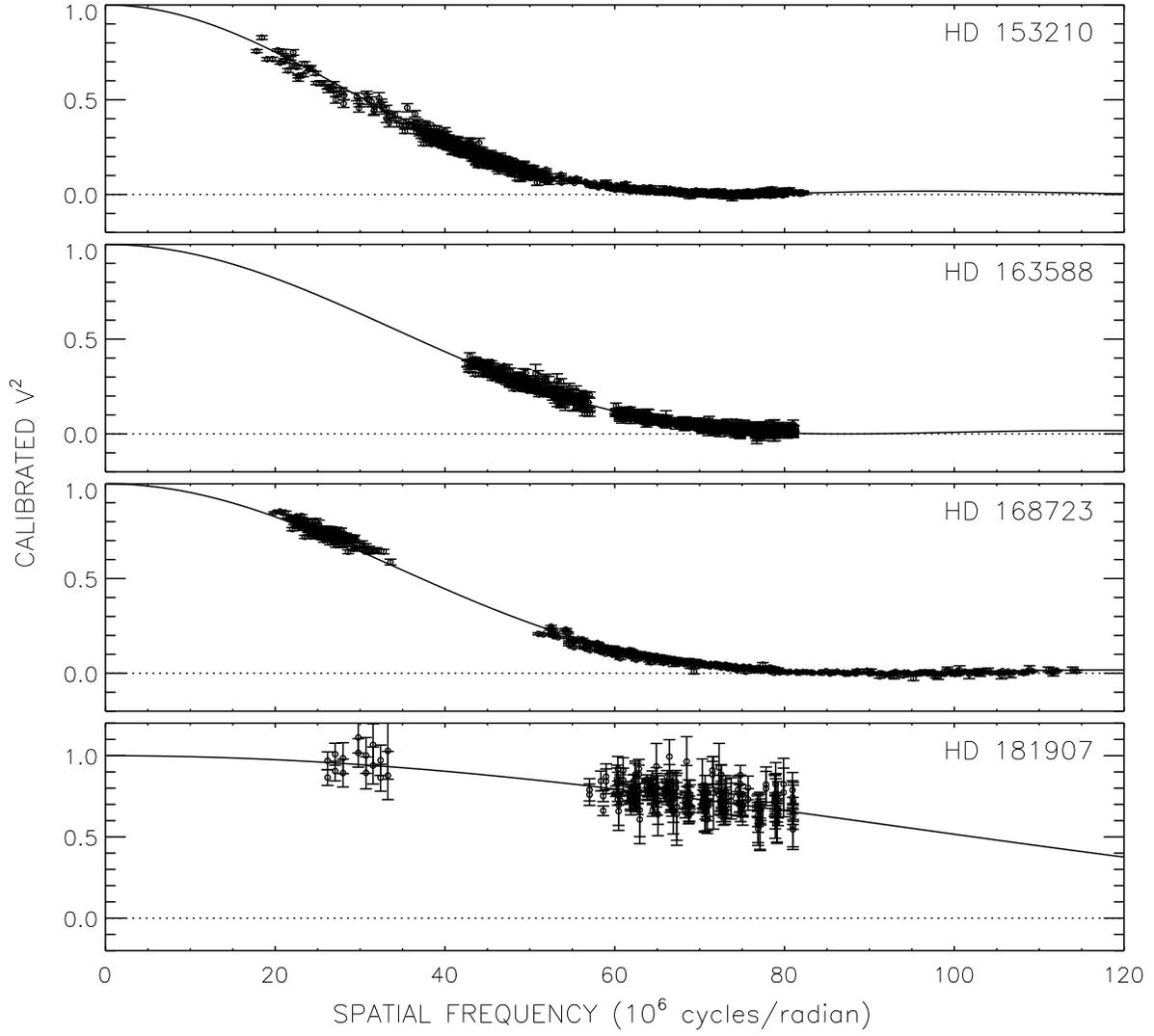}
\caption{$\theta_{\rm LD}$ fits for stars observed with two or three calibrators. The symbols are the same as in Figure \ref{ldplot1}.}
  \label{ldplot2}
\end{figure}

A more realistic model of a star's disk includes limb darkening (LD).  If a linear LD coefficient $\mu_\lambda$ is used,
\begin{equation}
V^2 = \left( {1-\mu_\lambda \over 2} + {\mu_\lambda \over 3} \right)^{-1}
\times
\left[(1-\mu_\lambda) {J_1(x_{\rm LD}) \over x_{\rm LD}} + \mu_\lambda {\left( \frac{\pi}{2} \right)^{1/2} \frac{J_{3/2}(x_{\rm LD})}{x_{\rm LD}^{3/2}}} \right] .
\end{equation}
where $x_{\rm LD} = \pi B\theta_{\rm LD}\lambda^{-1}$ \citep{1974MNRAS.167..475H}. We used effective temperature ($T_{\rm eff}$) and surface gravity (log $g$) values from the literature with a microturbulent velocity of 2 km s$^{\rm -1}$ and to obtain $\mu_\lambda$ from \citet{2011AandA...529A..75C}. These values and the resulting $\theta_{\rm LD}$ are listed in Table \ref{results}. Figures \ref{ldplot1} and \ref{ldplot2} show the $\theta_{\rm LD}$ fits for all the stars. The two stars with the largest percent uncertainties in the $\theta_{\rm LD}$ fit  (2$\%$) are HD 146791 and HD 181907. This is because their visibility curves are less well sampled with respect to spatial frequency than the other targets.

\begin{figure}[h]
\includegraphics[width=1.0\textwidth]{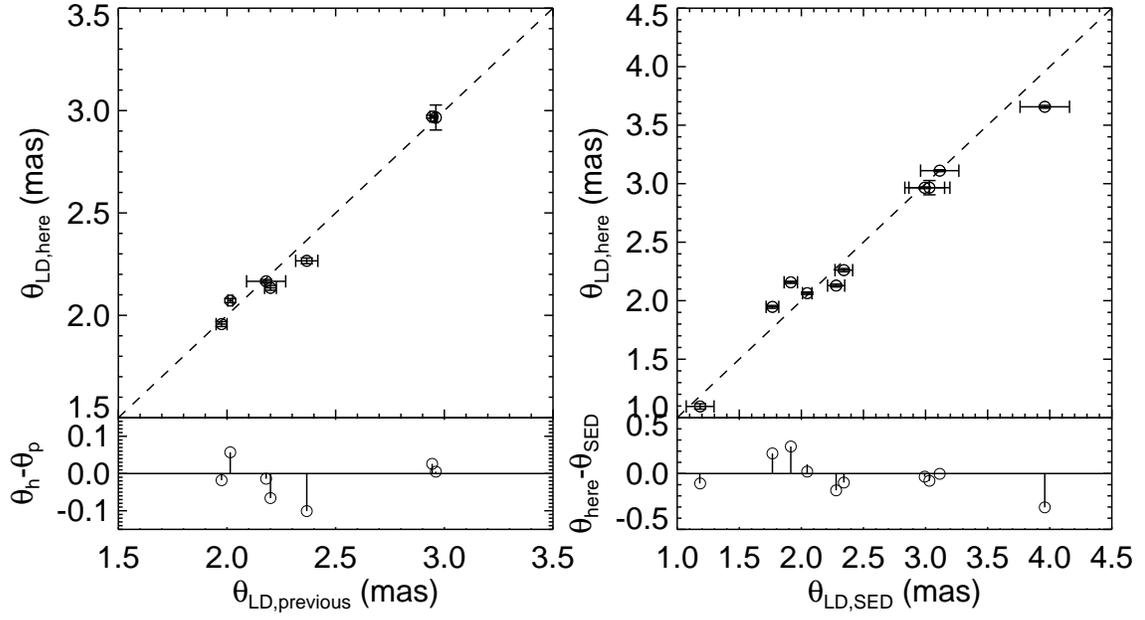}
\caption{Comparison between $\theta_{\rm LD}$ measured here and previous interferometric measurements from the literature (left panel) and compared to SED fits (right panel). The bottom panels show the residuals to the fit. The values used are listed in Table \ref{sed_lit}.}
  \label{diams}
\end{figure}

Seven of the 10 stars measured here had previous interferometric diameter measurements. They are listed in Table \ref{sed_lit} and are plotted against our values in Figure \ref{diams}. In all cases but one, the uncertainty on our diameter measurement is smaller than those from the literature, and they all agree to within 3-$\sigma$.

The uncertainty for the $\theta_{\rm LD}$ fit was derived using the method described in \citet{2010SPIE.7734E.103T}, who showed that a non-linear least-squares method does not sufficiently account for atmospheric effects on time scales shorter than the window between target and calibrator observations. They describe a bootstrap Monte Carlo method that treats the observations as groups of data points because the NPOI collects data in scans consisting of 16 channels simultaneously.\footnote{For every ``scan,'' 30 seconds of data are collected in each of the 16 wavelength channels with a measurement once every 2 milliseconds. During the processing described in Section 2, all the 30-second-scan's data points are averaged into one data point for each channel, so we go from 30 seconds of data per channel to one averaged data point per channel.} They discovered that when the 16 data points were analyzed individually, a single scan's deviation from the trend had a large impact on the resulting diameter and uncertainty calculation. On the other hand, when they preserved the inherent structure of the observational data by using the groups of 16 channels instead of individual data points, the uncertainty on the angular diameter was larger and more realistic. This method makes no assumptions about underlying uncertainties due to atmospheric effects, which are applicable to all stars observed using ground--based instruments. It should be noted that the number of calibrator stars used in the observations have no apparent effect on the $\theta_{\rm LD}$ fit uncertainty.

\subsection{Stellar Radius, Luminosity and Effective Temperature}

For each star, the parallax from \citet{2007AandA...474..653V} was converted into a distance, which we then combined with our measured $\theta_{\rm LD}$ to calculate the linear radius ($R$). In order to determine the luminosity ($L$) and $T_{\rm eff}$, we constructed each star's SED using the sources and technique of fitting spectral templates to observed photometry as described in Section 2. The resulting SED gave us the bolometric flux ($F_{\rm BOL}$) and allowed for the calculation of extinction $A_{\rm V}$ with the wavelength-dependent reddening relations of \citet{1989ApJ...345..245C}.

We combined our $F_{\rm BOL}$ values with the stars' distances to estimate $L$ using $L = 4 \pi d^2 F_{\rm BOL}$. We also combined the $F_{\rm BOL}$ with $\theta_{\rm LD}$ to determine each star's effective temperature by inverting the relation,
\begin{equation}
F_{\rm BOL} = {1 \over 4} \theta_{\rm LD}^2 \sigma T_{\rm eff}^4,
\end{equation}
where $\sigma$ is the Stefan-Boltzmann constant and $\theta_{\rm LD}$ is in radians. 

Because $\mu_\lambda$ is chosen based on a given $T_{\rm eff}$, we checked to see if $\mu_\lambda$ and therefore $\theta_{\rm LD}$ would change based on our new $T_{\rm eff}$. In most cases, $\mu_\lambda$ changed by 0.0 or 0.01, and the largest difference was 0.08 for HD 181907. The resulting $\theta_{\rm LD}$ values changed at most by 1$\%$, and all but three changed by 0.2$\%$ or less. This was well within the uncertainties on $\theta_{\rm LD}$, and re-calculating $T_{\rm eff}$ with the new $\theta_{\rm LD}$ made at most a 26 K difference (for HD 181907, which has an uncertainty of 199 K). These values all converged after this one iteration, and these are the final numbers listed in Table \ref{results}.


\section{Discussion}

Two scaling equations relate observed asteroseismic quantities to fundamental stellar parameters: 

\begin{equation}
\Delta \nu \propto M^{\frac{1}{2}} R^{- \frac{3}{2}},
\end{equation}

where $\Delta \nu$ is the large separation of oscillation modes of the same degree and consecutive orders and $M$ is the mass of the star \citep{1986ApJ...306L..37U}, and 

\begin{equation}
\nu_{\rm max} \propto M R^{-2} T_{\rm eff}^{- \frac{1}{2}},
\end{equation}

where $\nu_{\rm max}$ is the frequency of maximum oscillation power \citep{1991ApJ...368..599B, 1995AandA...293...87K}. These equations are often used to calculate stellar radii and masses from oscillation observations. However, when $R$ is measured interferometrically, we can test the relations themselves. 

We used $\Delta \nu$ and $\nu_{\rm max}$ from the references listed in Table \ref{compare} and assumed uncertainties of 1$\%$ in $\Delta \nu$ and 3$\%$ in $\nu_{\rm max}$ \citep{2012ApJ...760...32H} when no uncertainties were provided in the references. We combined the frequency measurements with effective temperatures from the literature ($T_{\rm lit}$) to calculate $R$ from the asteroseismic measurements alone. It should be noted that the $T_{\rm lit}$ used has little impact on $\nu_{\rm max}$. A variation of 100 K causes a 0.9$\%$ change in $\nu_{\rm max}$, which is typically on the order of or smaller than the uncertainties in $T_{\rm lit}$ \citep{2012ApJ...760...32H}.

\begin{figure}[h]
\includegraphics[width=0.6\textwidth]{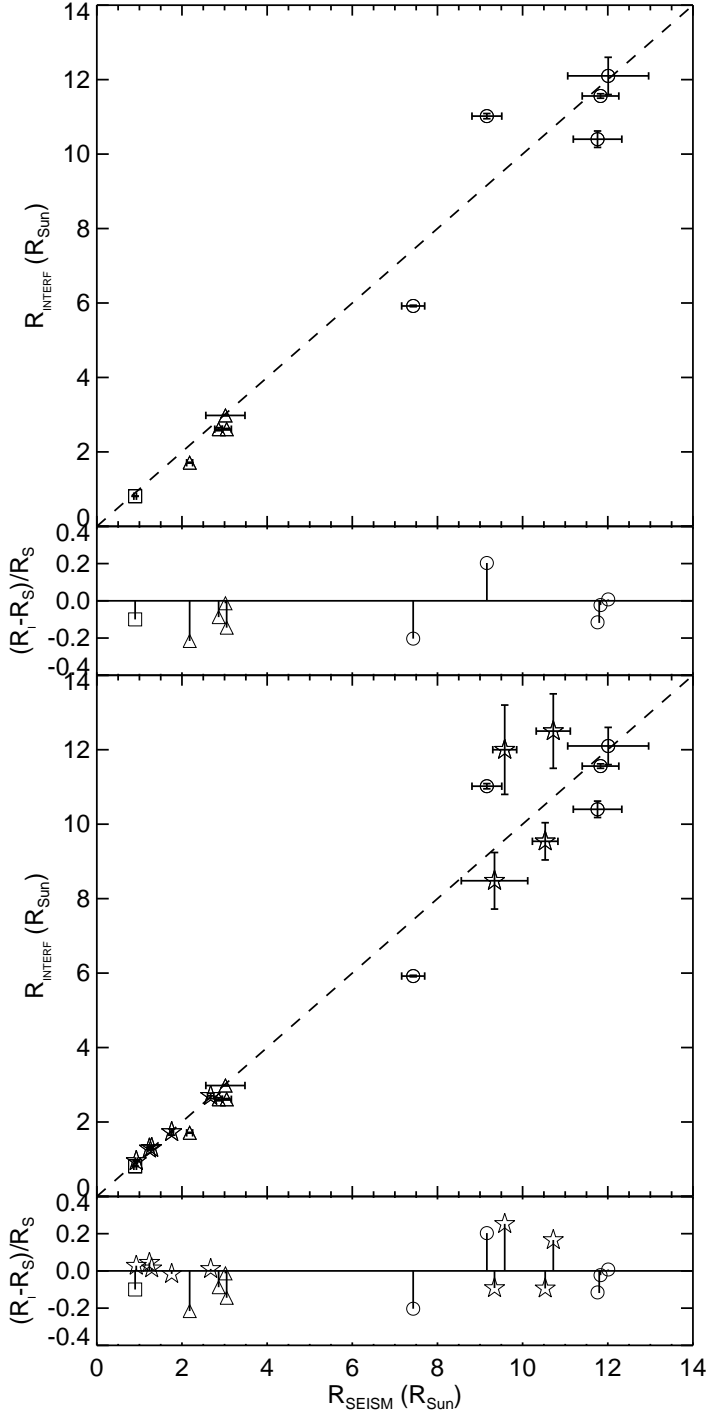}
\caption{Comparison between interferometrically measured radii and those determined asteroseismologically listed in Table \ref{compare}. The square represents the dwarf star, the triangles are subgiant stars, and the circles are giant stars. In the large bottom panel, the targets from \citet{2012ApJ...760...32H} are added in as stars. The dashed line is the 1:1 ratio. The small bottom panels show the residuals to the fit normalized to the asteroseismic radii.}
  \label{radii}
\end{figure}

Table \ref{compare} and Figure \ref{radii} show the results comparing the radii calculated from asteroseismology ($R_{\rm a}$) and those measured using interferometry ($R_{\rm i}$). The stars with the largest difference between the two are HD 153210, HD 161797, and HD 168723. For the latter two stars, previously published angular diameters agree with our measurement to 1$\%$ or less. HD 153210 has not been previously measured, and our diameter agrees with that predicted by the SED within 2-$\sigma$. For the remaining stars, $R_{\rm a}$ and $R_{\rm i}$ agree within 3-$\sigma$ ($\sim10\%$) in all cases except for HD 146791 and HD 150680, which agree within 4-$\sigma$ ($\sim15\%$). Three of the others (HD 163588, HD 181907, and HD 188512) agree within 1-$\sigma$ (1 to 2$\%$).

\citet{2004ESASP.559..113B} quote a range for $\nu_{\rm max}$ between 80 and 170 $\mu$Hz for HD 168723, and \citet{2009MNRAS.400L..80S} lists a value of 130 $\mu$Hz. If we use 130 $\mu$Hz in our calculation for $R_{\rm a}$, the result is 12.08$\pm$0.44 $R_\odot$, which is approximately twice the value of $R_{\rm i}$ = 5.92$\pm$0.02 $R_\odot$. However, if we use the lower end of the range, i.e., 80 $\mu$Hz, $R_{\rm a}$ is 7.43$\pm$0.27 $R_\odot$, which is still a 26$\%$ difference from the one presented here. We believe further asteroseismic observations of this star would be particularly interesting.

Some stars have more evenly and completely sampled data along the visibility curves than others; for example, HD 146791 and HD 181907 do not have as wide a range of measurements as a function of spatial frequency that other stars such as HD 121370 and HD 153210 display. We considered whether or not this would have an effect on the scatter in Figure \ref{radii} but the stars with sparsely sampled curves do not correspond to the outliers, so that is not the issue. In general, we do not observe any systematic trend as a function of size. The residual scatter in the giant stars is comparable to what \citet{2012ApJ...760...32H} found, and shows the relationships between observed $\Delta \nu$ and $\nu_{\rm max}$ and stellar radii are not as precise for evolved giant stars as they are for dwarf stars.

\begin{figure}[h]
\includegraphics[width=0.6\textwidth]{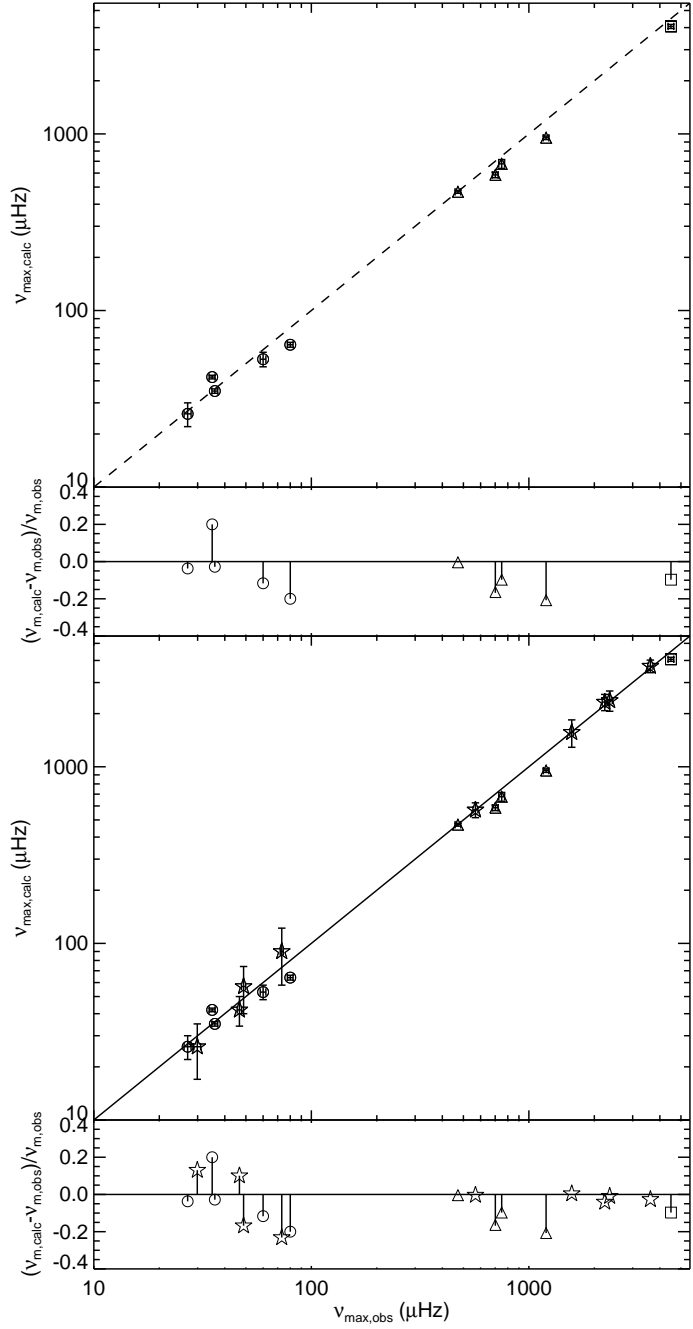}
\caption{Comparison between the calculated and observed $\nu_{\rm max}$ values from Table \ref{compare}. The symbols are the same as in Figure \ref{radii}.}
  \label{numax}
\end{figure}

The scaling relation for $\nu_{\rm max}$ is considered to be less robust than the relation for $\Delta \nu$ \citep{2012ApJ...760...32H}, so we wanted to test it. We combined Equation (3) with our $R_{\rm i}$ to calculate stellar masses, and then combined the masses with our new $T_{\rm eff}$ to calculate $\nu_{\rm max}$ values and compare them to the measured values. Table \ref{compare} lists and Figure \ref{numax} shows the results. The largest outliers are again HD 153210, HD 161797, and HD 168723 due to the discrepancies in calculated radii described above. HD 150680 also shows a 16$\%$ difference between the observed and calculated $\nu_{\rm max}$ value. Our angular diameter for this star matches those measured by \citet{2001AJ....122.2707N} and \citet{2003AJ....126.2502M} to within 3-$\sigma$, and it only differs from the SED estimate by 3$\%$. We note that the mass listed in \citet{2009ASPC..404..307K} is 1.19 $M_\odot$ while the mass determined by \citet{2001ESASP.464..431M} is 1.3 to 1.5 $M_\odot$. The latter agrees with the mass determined using our interferometric radius measurement: 1.33$\pm$0.04 $M_\odot$. In general we observe good agreement between the observed and calculated $\nu_{\rm max}$ within the uncertainty bars with no systematics with respect to stellar size or evolutionary status.



\section{Summary}

We measured the angular diameters of 10 stars using the NPOI. The combination of these observations with other information from the literature allowed us to calculate the stars' $R$, $T_{\rm eff}$, $F_{\rm BOL}$, and $L$. We compared our interferometric radius $R_{\rm i}$ values to those determined from asteroseismic scaling relations and found good agreement between the two, particularly for the less evolved stars. Then we also used $\Delta \nu$ from the literature and our $R_{\rm i}$ to calculate the stars' masses and $\nu_{\rm max}$ to put that scaling relation to the test as well. Again, the results agreed to within a few $\sigma$ in general. 

The relations work best for main-sequence stars and have limited precision for giant stars. Hopefully future observations with planned spacecraft such as \emph{Gaia} \citep{2003ASPC..298....3P} and \emph{TESS} \citep[Transiting Exoplanet Survey Satellite,][]{2009AAS...21430605R} as well as planned upgrades to existing interferometers such as the addition of longer baselines on the NPOI (increased resolution), new hardware (increased magnitude limit), and the eventual addition of large telescopes to the array \citep[see, e.g.,][]{Armstrong} will lead to significant improvements when combining data from two techniques.

\acknowledgments

The Navy Precision Optical Interferometer is a joint project of the Naval Research Laboratory and the U.S. Naval Observatory, in cooperation with Lowell Observatory, and is funded by the Office of Naval Research and the Oceanographer of the Navy. This research has made use of the SIMBAD database, operated at CDS, Strasbourg, France. This publication makes use of data products from the Two Micron All Sky Survey, which is a joint project of the University of Massachusetts and the Infrared Processing and Analysis Center/California Institute of Technology, funded by the National Aeronautics and Space Administration and the National Science Foundation.

\clearpage


\begin{deluxetable}{ccccccc}
\tablewidth{0pc}
\tablecaption{Observing Log and Calibrator Stars' Angular Diameters.\label{observations}}
\tablehead{
 \colhead{Target} & \colhead{Other} & \colhead{Calibrator} & \colhead{Date} & \colhead{Baselines}      & \colhead{$\#$} & \colhead{$\theta_{\rm LD,cal}$} \\
 \colhead{HD}     & \colhead{Name}  & \colhead{HD}         & \colhead{(UT)} & \colhead{Used$^\dagger$} & \colhead{Obs}  & \colhead{(mas)} \\ }
\startdata
10700  & $\tau$ Cet     & 11171  & 2005/09/17 & AC-W7, AE-W7        &  19 & 0.682$\pm$0.034 \\
       &                &        & 2005/09/18 & AC-W7               &   9 &                 \\
       &                &        & 2005/09/25 & AC-W7, AE-W7        &  44 &                 \\
121370 & $\eta$ Boo     & 122408 & 2012/05/10 & AW-E6, E6-W7        & 126 & 0.521$\pm$0.026 \\
       &                &        & 2012/05/11 & AE-AW, E6-W7        &  90 &                 \\
       &                &        & 2012/05/12 & E6-W7               &  27 &                 \\
       &                &        & 2012/05/15 & AW-E6, E6-W7        & 108 &                 \\
146791 & $\epsilon$ Oph & 141513 & 2010/03/21 & AW-W7               &  30 & 0.524$\pm$0.026 \\
       &                &        & 2010/03/28 & AW-W7               &  13 &                 \\
150680 & $\zeta$ Her    & 156164 & 2005/08/18 & AE-W7, E6-W7        &  50 & 0.887$\pm$0.044 \\
       &                &        & 2005/08/25 & AC-W7, AE-W7, E6-W7 & 120 &                 \\
153210 & $\kappa$ Oph   & 148112 & 2013/02/23 & AC-AE, AC-AW, AC-E6 &  18 & 0.443$\pm$0.022 \\
       &                &        & 2013/03/01 & AC-AW               &  12 &                 \\
       &                &        & 2013/03/02 & AC-E6, AW-E6        &  63 &                 \\
       &                &        & 2013/03/05 & AC-E6, AW-E6        & 198 &                 \\
       &                &        & 2013/03/06 & AC-E6, AW-E6        &  72 &                 \\
       &                &        & 2013/03/07 & AC-E6, AW-E6        &  90 &                 \\
       &                & 152614 & 2013/02/23 & AC-E6, AC-AW        &  34 & 0.316$\pm$0.016 \\
       &                &        & 2013/02/25 & AC-E6               &  14 &                 \\
       &                &        & 2013/03/02 & AC-E6, AW-E6        &  90 &                 \\
       &                &        & 2013/03/05 & AC-E6, AW-E6        & 261 &                 \\
       &                &        & 2013/03/07 & AC-E6, AW-E6        &  63 &                 \\
       &                &        & 2013/04/01 & AE-E6, AW-E6        &  53 &                 \\
       &                &        & 2013/04/02 & AE-E6, AW-E6        &  35 &                 \\
       &                &        & 2013/04/03 & AE-E6, AW-E6        &  54 &                 \\
       &                & 147547 & 2013/04/19 & AE-AW, AW-E6        &  68 & 0.970$\pm$0.049 \\
       &                &        & 2013/04/22 & AE-E6, AW-E6        & 115 &                 \\
       &                &        & 2013/04/23 & AE-E6, AW-E6        &  91 &                 \\
       &                &        & 2013/04/26 & AE-E6, AW-E6        & 165 &                 \\
161797 & $\mu$ Her      & 166014 & 2010/08/21 & AE-AW, AW-E6        & 270 & 0.596$\pm$0.030 \\
       &                &        & 2010/08/26 & AE-AW, AW-E6        &  58 &                 \\
163588 & $\xi$ Dra      & 159541 & 2013/04/27 & AE-AW, AW-E6        &  18 & 0.520$\pm$0.026 \\
       &                &        & 2013/04/28 & AW-E6               &  43 &                 \\
       &                &        & 2013/04/29 & AE-AW, AW-E6        &  97 &                 \\
       &                & 168151 & 2013/04/27 & AE-AW, AW-E6        &  43 & 0.655$\pm$0.033 \\
       &                &        & 2013/04/28 & AE-AW, AW-E6        &  49 &                 \\
       &                &        & 2013/04/29 & AE-AW, AW-E6        &  90 &                 \\
       &                &        & 2013/05/01 & AW-E6               &  10 &                 \\
       &                &        & 2013/05/03 & AE-AW               &  12 &                 \\
       &                &        & 2013/05/09 & AW-E6               & 170 &                 \\
       &                &        & 2013/05/12 & AW-E6               &  23 &                 \\
       &                &        & 2013/05/13 & AW-E6               & 236 &                 \\
       &                &        & 2013/05/14 & AE-AW               &  32 &                 \\
       &                & 184006 & 2013/05/01 & AW-E6               &  32 & 0.703$\pm$0.035 \\
       &                &        & 2013/05/12 & AE-AW, AW-E6        &  48 &                 \\
       &                &        & 2013/05/13 & AW-E6               & 104 &                 \\
       &                &        & 2013/05/14 & AE-AW               &  49 &                 \\
168723 & $\eta$ Ser     & 161868 & 2004/05/20 & AC-AE, AC-AW        &  90 & 0.668$\pm$0.033 \\
       &                &        & 2004/05/24 & AC-AE, AC-AW        &  17 &                 \\
       &                & 164353 & 2007/05/18 & AW-E6, E6-W7        & 120 & 0.371$\pm$0.022 \\
       &                &        & 2007/05/24 & AN-E6, AW-E6, E6-W7 & 278 &                 \\
       &                &        & 2007/05/25 & AN-E6, AW-E6        &  20 &                 \\
       &                &        & 2007/05/31 & AW-E6, E6-W7        & 128 &                 \\
181907 & HR 7349        & 177756 & 2013/05/21 & AW-E6               &  62 & 0.571$\pm$0.029 \\
       &                &        & 2013/05/31 & AC-AW, AW-E6        &  16 &                 \\
       &                &        & 2013/06/03 & AW-E6               &  23 &                 \\
       &                &        & 2013/06/05 & AW-E6               &  10 &                 \\
       &                & 184930 & 2013/05/21 & AW-E6               &  71 & 0.311$\pm$0.016 \\
       &                &        & 2013/05/31 & AC-AW, AW-E6        &  16 &                 \\
       &                &        & 2013/06/03 & AW-E6               &  21 &                 \\
       &                &        & 2013/06/05 & AW-E6               &  10 &                 \\
188512 & $\beta$ Aql    & 195810 & 2007/05/26 & AN-AW, AW-W7        &  14 & 0.394$\pm$0.020 \\
       &                &        & 2007/05/30 & AW-E6, E6-W7        &  57 &                 \\
       &                &        & 2007/05/31 & AN-E6, AW-E6, E6-W7 & 219 &                 \\
       &                &        & 2007/06/04 & AW-E6, E6-W7        & 140 &                 \\
       &                &        & 2007/06/09 & AN-E6, AW-E6, E6-W7 & 120 &                 \\
       &                &        & 2007/06/11 & AW-E6, E6-W7        &  89 &                 \\
\enddata
\tablecomments{$^\dagger$The maximum baseline lengths are AC-AE 18.9 m, AC-AW 22.2 m, AC-E6 34.4 m, AC-W7 51.3 m, AE-AW 37.5 m, AE-W7 64.2 m, AN-AW 38.2 m, AN-E6 45.6 m, AW-E6 53.3 m, AW-W7 29.5 m, and E6-W7 79.4 m. 
The $\theta_{\rm LD,cal}$ estimates were determined using the technique described in Section 2.}
\end{deluxetable}

\clearpage


\begin{deluxetable}{lcccccccrcclc}
\rotate
\tablewidth{0pc}
\tabletypesize{\scriptsize}
\tablecaption{Stellar Parameters. \label{results}}

\tablehead{\colhead{Target} & \colhead{Spectral} & \colhead{Parallax} & \colhead{ }                   & \colhead{$\theta_{\rm UD}$} & \colhead{$\theta_{\rm LD}$} & \colhead{$\sigma_{\rm LD}$} & \colhead{$\#$} & \colhead{$R_{\rm linear}$} & \colhead{$L$}   & \colhead{$F_{\rm BOL}$}                      & \colhead{$T_{\rm eff}$} & \colhead{$\sigma_{\rm Teff}$} \\ 
           \colhead{HD}     & \colhead{Type}     & \colhead{(mas)}    & \colhead{$\mu_{\rm \lambda}$} & \colhead{(mas)}             & \colhead{(mas)}             & \colhead{($\%$)}            & \colhead{Cals} & \colhead{($R_\odot$)}      & \colhead{($L_{\odot}$)} & \colhead{(10$^{-8}$ erg s$^{-1}$ cm$^{-2}$)} & \colhead{(K)}           & \colhead{$\%$} }
\startdata
10700  & G8.5 V    & 273.96$\pm$0.17 & 0.60 & 1.952$\pm$0.003 & 2.072$\pm$0.010 & 0.5 & 1 &  0.81$\pm$0.01 &  0.47$\pm$0.01 & 113.0$\pm$0.3 & 5301$\pm$13  & 0.2 \\
121370 & G0 IV     &  87.75$\pm$1.24 & 0.52 & 2.023$\pm$0.002 & 2.134$\pm$0.012 & 0.6 & 1 &  2.61$\pm$0.04 &  8.69$\pm$0.25 & 214.0$\pm$0.7 & 6128$\pm$18  & 0.3 \\
146791 & G9.5 III  &  30.64$\pm$0.20 & 0.65 & 2.772$\pm$0.007 & 2.966$\pm$0.061 & 2.1 & 1 & 10.40$\pm$0.22 & 56.64$\pm$1.29 & 170.0$\pm$3.2 & 4907$\pm$55  & 1.1 \\
150680 & G0 IV     &  93.32$\pm$0.47 & 0.53 & 2.175$\pm$0.001 & 2.266$\pm$0.014 & 0.6 & 1 &  2.61$\pm$0.02 &  8.12$\pm$0.08 & 226.0$\pm$0.6 & 6029$\pm$19  & 0.3 \\
153210 & K2 III    &  35.66$\pm$0.20 & 0.71 & 3.479$\pm$0.001 & 3.657$\pm$0.013 & 0.4 & 3 & 11.02$\pm$0.07 & 50.67$\pm$1.14 & 206.0$\pm$4.0 & 4367$\pm$24  & 0.5 \\
161797 & G5 IV     & 123.33$\pm$0.16 & 0.58 & 1.851$\pm$0.002 & 1.957$\pm$0.012 & 0.6 & 1 &  1.71$\pm$0.01 &  2.10$\pm$0.01 & 102.0$\pm$0.2 & 5317$\pm$16  & 0.3 \\
163588 & K2 III    &  28.98$\pm$0.12 & 0.70 & 2.894$\pm$0.002 & 3.116$\pm$0.008 & 0.3 & 3 & 11.56$\pm$0.06 & 47.30$\pm$0.44 & 127.0$\pm$0.6 & 4451$\pm$7   & 0.2 \\
168723 & K0 III-IV &  53.93$\pm$0.18 & 0.65 & 2.852$\pm$0.001 & 2.970$\pm$0.007 & 0.2 & 2 &  5.92$\pm$0.02 & 17.85$\pm$0.13 & 166.0$\pm$0.4 & 4875$\pm$7   & 0.1 \\
181907 & G8 III    &   9.67$\pm$0.34 & 0.67 & 1.038$\pm$0.009 & 1.089$\pm$0.023 & 2.1 & 2 & 12.10$\pm$0.50 & 96.68$\pm$16.0 &  29.5$\pm$4.3 & 5227$\pm$199 & 3.8 \\
188512 & G9.5 IV   &  78.00$\pm$0.20 & 0.63 & 2.042$\pm$0.001 & 2.166$\pm$0.009 & 0.4 & 1 &  2.98$\pm$0.01 &  4.99$\pm$0.03 &  97.1$\pm$0.2 & 4992$\pm$11  & 0.2 \\
\enddata
\tablecomments{The parallaxes are from \citet{2007AandA...474..653V}; the $\mu_{\rm \lambda}$ coefficients are from \citet{2011AandA...529A..75C} in the $R$-band with a microturbulent velocity of 2 km s$^{\rm -1}$. The sources of $T_{\rm eff}$ and log $g$ used to determine $\mu_{\rm \lambda}$ were the following: \citet{2011AandA...531A.165P} for HD 10700, HD 121370, HD 161797, HD 168723, and HD 188512; \citet{2011AandA...525A..71W} for HD 146791 and HD 163588; \citet{2007astro.ph..3658P} for HD 150680; \citet{2012MNRAS.419L..34M} for HD 153210; and \citet{2006ApJ...638.1004A} for HD 181907. }
\end{deluxetable}

\clearpage


\begin{deluxetable}{lcccclc}
\tablewidth{0pc}
\tablecaption{Angular Diameter Comparison. \label{sed_lit}}

\tablehead{\colhead{Target} & \colhead{$\theta_{\rm LD,this work}$} & \colhead{$\theta_{\rm LD,SED}$} & \colhead{$\%$} & \colhead{$\theta_{\rm LD,previous}$} & \colhead{ }         & \colhead{$\%$} \\ 
           \colhead{HD}     & \colhead{(mas)}                       & \colhead{(mas)}                 & \colhead{diff} & \colhead{(mas)}                      & \colhead{Reference} & \colhead{diff} } 
\startdata
10700  & 2.072$\pm$0.010 & 2.047$\pm$0.038 & 1   &  1.97$\pm$0.05  & \citet{2003AandA...406L..15P} & 5 \\
       &                 &                 &     & 2.015$\pm$0.011 & \citet{2007AandA...475..243D} & 3 \\
121370 & 2.134$\pm$0.012 & 2.280$\pm$0.069 & 7   &  2.28$\pm$0.07  & \citet{2001AJ....122.2707N}   & 7 \\
       &                 &                 &     & 2.269$\pm$0.025 & \citet{2003AJ....126.2502M}   & 6 \\
       &                 &                 &     & 2.200$\pm$0.027 & \citet{2005AandA...436..253T} & 3 \\
146791 & 2.966$\pm$0.061 & 3.031$\pm$0.165 & 2   & 2.961$\pm$0.007 & \citet{2009AandA...503..521M} & 0.2 \\
150680 & 2.266$\pm$0.014 & 2.342$\pm$0.071 & 3   &  2.49$\pm$0.09  & \citet{2001AJ....122.2707N}   & 10 \\
       &                 &                 &     & 2.367$\pm$0.051 & \citet{2003AJ....126.2502M}   & 4 \\
153210 & 3.657$\pm$0.013 & 3.960$\pm$0.199 & 8   & N/A             & N/A                           & -- \\
161797 & 1.957$\pm$0.012 & 1.767$\pm$0.051 & 10  & 1.953$\pm$0.039 & \citet{2003AJ....126.2502M}   & 0.2 \\
       &                 &                 &     & 1.975$\pm$0.025 & \citet{2013AandA...555A.104A} & 1 \\
163588 & 3.116$\pm$0.008 & 3.114$\pm$0.154 & 0.1 & N/A             & N/A                           & -- \\
168723 & 2.970$\pm$0.007 & 2.993$\pm$0.160 & 1   & 2.944$\pm$0.010 & \citet{2010AandA...517A..64M} & 1 \\
181907 & 1.089$\pm$0.023 & 1.185$\pm$0.112 & 8   & N/A             & N/A                           & -- \\
188512 & 2.166$\pm$0.009 & 1.915$\pm$0.054 & 12  &  2.18$\pm$0.09  & \citet{1999AJ....118.3032N}   & 1 \\
\enddata
\tablecomments{The SED fit was created using the method described in Section 3.2. If more than one diameter is available in the literature, we used the most recent one when plotting the results in Figure \ref{diams}.}
\end{deluxetable}

\clearpage


\begin{deluxetable}{cccccccccccc}
\rotate
\tablewidth{0pc}
\tabletypesize{\scriptsize}
\tablecaption{Comparing Interferometric and Asteroseismic Results. \label{compare}}

\tablehead{\colhead{Target} & \colhead{$T_{\rm i}$} & \colhead{$T_{\rm lit}$} & \colhead{$R_{\rm i}$} & \colhead{$R_{\rm a}$} & \colhead{$\%$} & \colhead{$\Delta \nu$} & \colhead{$\nu_{\rm max}$} & \colhead{ }   & \colhead{$M$}         & \colhead{$\nu_{\rm max,calc}$} & \colhead{$\%$}  \\ 
           \colhead{HD}     & \colhead{(K)}         & \colhead{(K)}           & \colhead{($R_\odot$)} & \colhead{($R_\odot$)} & \colhead{diff} & \colhead{($\mu$Hz)}    & \colhead{($\mu$Hz)}       & \colhead{Reference} & \colhead{($M_\odot$)} & \colhead{($\mu$Hz)}            & \colhead{diff}  }
\startdata
10700  & 5301$\pm$13  & 5348$\pm$45  &  0.81$\pm$0.01 &  0.90$\pm$0.03 & 12 & 169            & 4500           & \citet{2009AandA...494..237T} & 0.84$\pm$0.02 & 4062$\pm$108 & 10  \\
121370 & 6128$\pm$18  & 5967$\pm$45  &  2.61$\pm$0.04 &  2.86$\pm$0.10 &  9 & 39.9$\pm$0.1   &  750           & \citet{2005AandA...434.1085C} & 1.56$\pm$0.07 &  677$\pm$37  & 10  \\
146791 & 4907$\pm$55  & 4918$\pm$28  & 10.40$\pm$0.22 & 11.76$\pm$0.43 & 13 &  5.3$\pm$0.1   &   60           & \citet{2007AandA...468.1033B} & 1.74$\pm$0.13 &   53$\pm$5   & 11  \\
150680 & 6029$\pm$19  & 5758$\pm$81  &  2.61$\pm$0.02 &  3.05$\pm$0.11 & 17 & 37.01          &  701$^{\rm a}$ & \citet{2009ASPC..404..307K}   & 1.34$\pm$0.04 &  586$\pm$21  & 16  \\
153210 & 4637$\pm$24  & 4559$\pm$116 & 11.02$\pm$0.07 &  9.16$\pm$0.35 & 16 &  4.5           &   35           & \citet{2009MNRAS.400L..80S}   & 1.49$\pm$0.04 &   42$\pm$1   & 19  \\
161797 & 5317$\pm$16  & 5454$\pm$35  &  1.71$\pm$0.01 &  2.18$\pm$0.08 & 28 & 56.5$\pm$0.07  & 1200           & \citet{2008ApJ...676.1248B}   & 0.87$\pm$0.02 &  951$\pm$22  & 21  \\
163588 & 4451$\pm$7   & 4483$\pm$25  & 11.56$\pm$0.06 & 11.83$\pm$0.43 &  2 &  4             &   36           & \citet{2009MNRAS.400L..80S}   & 1.36$\pm$0.03 &   35$\pm$1   &  2  \\
168723 & 4875$\pm$7   & 4923$\pm$63  &  5.92$\pm$0.02 &  7.43$\pm$0.27 & 26 &  7.7           &   80           & \citet{2004ESASP.559..113B}   & 0.68$\pm$0.02 &   64$\pm$2   & 20  \\
181907 & 5227$\pm$199 & 5637$\pm$228 & 12.10$\pm$0.50 & 12.01$\pm$0.52 &  1 &  3.47$\pm$0.12 &   27$^{\rm b}$ & \citet{2010AandA...509A..73C} & 1.17$\pm$0.17 &   26$\pm$4   &  5  \\
188512 & 4992$\pm$11  & 5082$\pm$69  &  2.98$\pm$0.01 &  3.02$\pm$0.11 &  1 & 29.56$\pm$0.10 &  472$\pm$72    & \citet{2012AandA...537A...9C} & 1.28$\pm$0.02 &  470$\pm$9  & 0.4  \\
\enddata
\tablecomments{$T_{\rm i}$ is the interferometrically calculated effective temperature from Table \ref{results}; $T_{\rm lit}$ is the effective temperature from the literature as listed in Table \ref{results}. $R_{\rm i}$ is the interferometrically measured radius; $R_{\rm a}$ is the radius calculated using $\nu_{\rm max}$ and $\Delta \nu$ from asteroseismic observations as well as $T_{\rm lit}$; $\nu_{\rm max}$ and $\Delta \nu$ are from the references listed; $M$ is the mass calculated using $\Delta \nu$ and $R_{\rm i}$; and $\nu_{\rm max,calc}$ is the $\nu_{\rm max}$ calculated using $M$ and $T_{\rm i}$. \\
$^{\rm a}$No $\nu_{\rm max}$ was listed, so it was calculated using $T_{\rm eff}$ from \citet{2001ESASP.464..431M} and $M$ and $R$ from \citet{2009ASPC..404..307K}. \\
$^{\rm b}$No $\nu_{\rm max}$ was listed, so it was calculated using $T_{\rm eff}$, $M$, and $R$ from \citet{2010AandA...509A..73C}.
}
\end{deluxetable}

\clearpage


\clearpage

\clearpage

\clearpage

\clearpage

\begin{thebibliography}{}

\bibitem[Absil et al.(2013)]{2013AandA...555A.104A} Absil, O., Defr{\`e}re, D., Coud{\'e} du Foresto, V., et al.\ 2013, \aap, 555, A104 

\bibitem[Ammons et al.(2006)]{2006ApJ...638.1004A} Ammons, S.~M., Robinson, S.~E., Strader, J., et al.\ 2006, \apj, 638, 1004 

\bibitem[Armstrong et al.(1998)]{1998ApJ...496..550A} Armstrong, J.~T.,  Mozurkewich, D., Rickard, L.~J, et al.\ 1998, \apj, 496, 550 

\bibitem[Armstrong et al.(2013)]{Armstrong} Armstrong, J.~T.,  Hutter, D. J., Baines, E. K., et al.\ 2013, Journal of Astronomical Instrumentation, in press

\bibitem[Auvergne et al.(2009)]{2009AandA...506..411A} Auvergne, M., Bodin, P., Boisnard, L., et al.\ 2009, \aap, 506

\bibitem[Baglin et al.(2006)]{2006ESASP.624E..34B} Baglin, A., Michel, E., Auvergne, M., \& COROT Team 2006, Proceedings of SOHO 18/GONG 2006/HELAS I, Beyond the spherical Sun, 624

\bibitem[Baines et al.(2012)]{2012ApJ...761...57B} Baines, E.~K., White, R.~J., Huber, D., et al.\ 2012, \apj, 761, 57 

\bibitem[Barban et al.(2004)]{2004ESASP.559..113B} Barban, C., De Ridder, J., Mazumdar, A., et al.\ 2004, SOHO 14 Helio- and Asteroseismology: Towards a Golden Future, 559, 113 

\bibitem[Barban et al.(2007)]{2007AandA...468.1033B} Barban, C., Matthews, J.~M., De Ridder, J., et al.\ 2007, \aap, 468, 1033 

\bibitem[Bonanno et al.(2008)]{2008ApJ...676.1248B} Bonanno, A., Benatti, S., Claudi, R., et al.\ 2008, \apj, 676, 1248 

\bibitem[Borucki et al.(2010)]{2010Sci...327..977B} Borucki, W.~J., Koch, D., Basri, G., et al.\ 2010, Science, 327, 977 

\bibitem[Brown et al.(1991)]{1991ApJ...368..599B} Brown, T.~M., Gilliland, R.~L., Noyes, R.~W., \& Ramsey, L.~W.\ 1991, \apj, 368, 599 

\bibitem[Brown \& Gilliland(1994)]{1994ARAandA..32...37B} Brown, T.~M., \& Gilliland, R.~L.\ 1994, \araa, 32, 37 

\bibitem[Cardelli et al.(1989)]{1989ApJ...345..245C} Cardelli, J.~A., Clayton, G.~C., \& Mathis, J.~S.\ 1989, \apj, 345, 245 

\bibitem[Carrier et al.(2005)]{2005AandA...434.1085C} Carrier, F., Eggenberger, P., \& Bouchy, F.\ 2005, \aap, 434, 1085 

\bibitem[Carrier et al.(2010)]{2010AandA...509A..73C} Carrier, F., De Ridder, J., Baudin, F., et al.\ 2010, \aap, 509, A73 

\bibitem[Chaplin et al.(2011)]{2011Sci...332..213C} Chaplin, W.~J., Kjeldsen, H., Christensen-Dalsgaard, J., et al.\ 2011, Science, 332, 213 

\bibitem[Christensen-Dalsgaard(2004)]{2004SoPh..220..137C} Christensen-Dalsgaard, J.\ 2004, \solphys, 220, 137 

\bibitem[Claret \& Bloemen(2011)]{2011AandA...529A..75C} Claret, A., \& Bloemen, S.\ 2011, \aap, 529, A75

\bibitem[Corsaro et al.(2012)]{2012AandA...537A...9C} Corsaro, E., Grundahl, F., Leccia, S., et al.\ 2012, \aap, 537, A9 

\bibitem[Cunha et al.(2007)]{2007AandARv..14..217C} Cunha, M.~S., Aerts, C., Christensen-Dalsgaard, J., et al.\ 2007, \aapr, 14, 217 

\bibitem[Cutri et al.(2003)]{2003tmc..book.....C} Cutri, R.~M., et al.\ 2003, The IRSA 2MASS All-Sky Point Source Catalog, NASA/IPAC Infrared Science Archive

\bibitem[di Folco et al.(2007)]{2007AandA...475..243D} di Folco, E., Absil, O., Augereau, J.-C., et al.\ 2007, \aap, 475, 243 

\bibitem[Eggen(1968)]{1968tcpn.book.....E} Eggen, O.~J.\ 1968, London, H.M.S.O., 1968

\bibitem[Gezari et al.(1993)]{1993cio..book.....G} Gezari, D.~Y., Schmitz, M., Pitts, P.~S., \& Mead, J.~M.\ 1993, Unknown  

\bibitem[Glushneva et al.(1983)]{1983TrSht..53...50G} Glushneva, I.~N., Doroshenko, V.~T., Fetisova, T.~S., et al.\ 1983, Trudy Gosudarstvennogo Astronomicheskogo Instituta, 53, 50 

\bibitem[Glushneva et al.(1998)]{1998yCat.3207....0G} Glushneva, I.~N., Doroshenko, V.~T., Fetisova, T.~S., et al.\ 1998, VizieR Online Data Catalog, 3207, 0 

\bibitem[Golay(1972)]{1972VA.....14...13G} Golay, M.\ 1972, Vistas in Astronomy, 14, 13 

\bibitem[H{\"a}ggkvist \& Oja(1970)]{1970AandAS....1..199H} H{\"a}ggkvist, L., \& Oja, T.\ 1970, \aaps, 1, 199

\bibitem[Hanbury Brown et al.(1974)]{1974MNRAS.167..475H} Hanbury Brown, R., Davis, J., Lake, R.~J.~W., \& Thompson, R.~J.\ 1974, \mnras, 167, 475

\bibitem[Huber et al.(2012a)]{2012MNRAS.423L..16H} Huber, D., Ireland, M.~J., Bedding, T.~R., et al.\ 2012a, \mnras, 423, L16 

\bibitem[Huber et al.(2012b)]{2012ApJ...760...32H} Huber, D., Ireland, M.~J., Bedding, T.~R., et al.\ 2012b, \apj, 760, 32 

\bibitem[Hummel et al.(2003)]{2003AJ....125.2630H} Hummel, C.~A., Benson, J.~A., Hutter, D.~J., et al.\ 2003, \aj, 125, 2630 

\bibitem[Jasevicius et al.(1990)]{1990VilOB..85...50J} Jasevicius, V., Kuriliene, G., Strazdaite, V., et al.\ 1990, Vilnius Astronomijos Observatorijos Biuletenis, 85, 50 

\bibitem[Johnson et al.(1966)]{1966CoLPL...4...99J} Johnson, H.~L., Mitchell, R.~I., Iriarte, B., \& Wisniewski, W.~Z.\ 1966, Communications of the Lunar and Planetary Laboratory, 4, 99 

\bibitem[Kallinger et al.(2009)]{2009ASPC..404..307K} Kallinger, T., Weiss, W.~W., De Ridder, J., Hekker, S., \& Barban, C.\ 2009, The Eighth Pacific Rim Conference on Stellar Astrophysics: A Tribute to Kam-Ching Leung, 404, 307 

\bibitem[Kharitonov et al.(1997)]{1997yCat.3202....0K} Kharitonov, A.~V., Tereshchenko, V.~M., \& Knyazeva, L.~N.\ 1997, VizieR Online Data Catalog, 3202, 0 

\bibitem[Kjeldsen \& Bedding(1995)]{1995AandA...293...87K} Kjeldsen, H., \& Bedding, T.~R.\ 1995, \aap, 293, 87 

\bibitem[Koch et al.(2010)]{2010ApJ...713L..79K} Koch, D.~G., Borucki, W.~J., Basri, G., et al.\ 2010, \apjl, 713, L79 

\bibitem[Kornilov et al.(1991)]{1991TrSht..63....1K} Kornilov, V.~G., Volkov, I.~M., Zakharov, A.~I., et al.\ 1991, Trudy Gosudarstvennogo Astronomicheskogo Instituta, 63, 1 

\bibitem[Ljunggren \& Oja(1965)]{1965ArA.....3..439L} Ljunggren, B., \& Oja, T.\ 1965, Arkiv for Astronomi, 3, 439 

\bibitem[Marti{\'c} et al.(2001)]{2001ESASP.464..431M} Marti{\'c}, M., Lebrun, J.~C., Schmitt, J., Appourchaux, T., \& Bertaux, J.~L.\ 2001, SOHO 10/GONG 2000 Workshop: Helio- and Asteroseismology at the Dawn of the Millennium, 464, 431 

\bibitem[Mazumdar et al.(2009)]{2009AandA...503..521M} Mazumdar, A., M{\'e}rand, A., Demarque, P., et al.\ 2009, \aap, 503, 521 

\bibitem[McClure \& Forrester(1981)]{1981PDAO...15..439M} McClure, R.~D., \& Forrester, W.~T.\ 1981, Publications of the Dominion Astrophysical Observatory Victoria, 15, 439 

\bibitem[M{\'e}rand et al.(2010)]{2010AandA...517A..64M} M{\'e}rand, A., Kervella, P., Barban, C., et al.\ 2010, \aap, 517, A64 

\bibitem[Mermilliod et al.(1997)]{1997AandAS..124..349M} Mermilliod, J.-C., Mermilliod, M., \& Hauck, B.\ 1997, \aaps, 124, 349 

\bibitem[Morel \& Miglio(2012)]{2012MNRAS.419L..34M} Morel, T., \& Miglio, A.\ 2012, \mnras, 419, L34 

\bibitem[Mozurkewich et al.(2003)]{2003AJ....126.2502M} Mozurkewich, D., Armstrong, J.~T., Hindsley, R.~B., et al.\ 2003, \aj, 126, 2502 

\bibitem[Nordgren et al.(1999)]{1999AJ....118.3032N} Nordgren, T.~E., Germain, M.~E., Benson, J.~A., et al.\ 1999, \aj, 118, 3032 

\bibitem[Nordgren et al.(2001)]{2001AJ....122.2707N} Nordgren, T.~E., Sudol, J.~J., \& Mozurkewich, D.\ 2001, \aj, 122, 2707 

\bibitem[Olsen(1993)]{1993AandAS..102...89O} Olsen, E.~H.\ 1993, \aaps, 102, 89 

\bibitem[Perryman(2003)]{2003ASPC..298....3P} Perryman, M.~A.~C.\ 2003, GAIA Spectroscopy: Science and Technology, 298, 3 

\bibitem[Pickles(1998)]{1998PASP..110..863P} Pickles, A.~J.\ 1998, \pasp, 110, 863

\bibitem[Pijpers et al.(2003)]{2003AandA...406L..15P} Pijpers, F.~P., Teixeira, T.~C., Garcia, P.~J., et al.\ 2003, \aap, 406, L15 

\bibitem[Press et al.(1992)]{1992nrca.book.....P} Press, W.~H., Teukolsky, S.~A., Vetterling, W.~T., \& Flannery, B.~P.\ 1992, Numerical recipes in C. The art of scientific computing (Cambridge: University Press, c1992, 2nd ed.)

\bibitem[Prugniel et al.(2007)]{2007astro.ph..3658P} Prugniel, P., Soubiran, C., Koleva, M., \& Le Borgne, D.\ 2007, arXiv:astro-ph/0703658 

\bibitem[Prugniel et al.(2011)]{2011AandA...531A.165P} Prugniel, P., Vauglin, I., \& Koleva, M.\ 2011, \aap, 531, A165 

\bibitem[Ricker et al.(2009)]{2009AAS...21430605R} Ricker, G.~R., Latham, D.~W., Vanderspek, R.~K., et al.\ 2009, American Astronomical Society Meeting Abstracts \#214, 214, \#306.05 

\bibitem[Shao \& Colavita(1992)]{1992ARAandA..30..457S} Shao, M., \& Colavita, M.~M.\ 1992, \araa, 30, 457 

\bibitem[Stello et al.(2009)]{2009MNRAS.400L..80S} Stello, D., Chaplin, W.~J., Basu, S., Elsworth, Y., \& Bedding, T.~R.\ 2009, \mnras, 400, L80 

\bibitem[Teixeira et al.(2009)]{2009AandA...494..237T} Teixeira, T.~C., Kjeldsen, H., Bedding, T.~R., et al.\ 2009, \aap, 494, 237 

\bibitem[Th{\'e}venin et al.(2005)]{2005AandA...436..253T} Th{\'e}venin, F., Kervella, P., Pichon, B., et al.\ 2005, \aap, 436, 253 

\bibitem[Tycner et al.(2010)]{2010SPIE.7734E.103T} Tycner, C., Hutter, D.~J., \& Zavala, R.~T.\ 2010, \procspie, 7734, 103T

\bibitem[Ulrich(1986)]{1986ApJ...306L..37U} Ulrich, R.~K.\ 1986, \apjl, 306, L37 

\bibitem[van Leeuwen(2007)]{2007AandA...474..653V} van Leeuwen, F.\ 2007, \aap, 474, 653 

\bibitem[Walker et al.(2003)]{2003PASP..115.1023W} Walker, G., Matthews, J., Kuschnig, R., et al.\ 2003, \pasp, 115, 1023 

\bibitem[Wall \& Jenkins(2003)]{2003psa..book.....W} Wall, J.~V., \& Jenkins, C.~R.\ 2003, Practical Statistics for Astronomers (Princeton Series in Astrophysics)

\bibitem[Wu et al.(2011)]{2011AandA...525A..71W} Wu, Y., Singh, H.~P., Prugniel, P., Gupta, R., \& Koleva, M.\ 2011, \aap, 525, A71 

\end{thebibliography}
\end{document}